\documentclass[reprint,prd,superscriptaddress,showpacs,nofootinbib,floatfix,preprintnumbers]{revtex4-1}
\pdfoutput=1
\usepackage{graphicx}
\usepackage{epsfig}
\usepackage{amsmath,amssymb,hyperref,enumerate}
\usepackage{pstricks}
\usepackage{slashed}
\usepackage{amsfonts}
\usepackage{url}
\usepackage{color}
\usepackage{appendix}
\usepackage{bm}
\usepackage{subfig}
\hypersetup{colorlinks,citecolor= black,linkcolor= black}
\definecolor{nicered}{rgb}{0.7,0.1,0.1}
\definecolor{nicegreen}{rgb}{0.1,0.5,0.1}

\def\({\left(}
\def\){\right)}
\def\[{\left[}
\def\]{\right]}

\begin{document}
\preprint{MAN/HEP/2015/07, UMD-PP-015-008}

\title{A TeV Scale Model for Baryon and Lepton Number Violation \\ and Resonant Baryogenesis}

\author{P. S. Bhupal Dev}
\affiliation{Consortium for Fundamental Physics, School of Physics and Astronomy, University of Manchester, Manchester M13 9PL, United Kingdom}
 
\author{Rabindra N. Mohapatra}
\affiliation{Maryland Center for Fundamental Physics and Department of Physics,
University of Maryland, College Park, Maryland 20742, USA}

\begin{abstract} A simple TeV scale model for baryon and lepton number violation is presented, where neutrino mass arises via a one loop radiative seesaw effect  and $B$-violation obeys $\Delta B=2$ selection rule. The stability of proton is connected to the neutrino mass generation. Matter-antimatter asymmetry is generated in this model via 
resonant baryogenesis mechanism. 
\end{abstract}
\maketitle
\section{Introduction}\label{sec:1}
Baryon number ($B$) violation is an essential requirement for understanding the origin of matter in the Universe according to Sakharov's criteria~\cite{sakharov}. 
Some relevant questions that arise are: (i)  what are the selection rules for $B$-violation; (ii) what is the scale of $B$-violation; and (iii) what is the physics associated them? At the level of effective theories, the various Standard Model (SM)-invariant operators that lead to $B$-violation can be classified according to their mass dimensions~\cite{weinberg} and they give a rough idea about the mass scales of their couplings which make them observable. This provides a way to probe very high scales using only low-energy experiments. For instance, 
a typical effective leading dimension-6 operator is  $QQQL$ which leads to proton decay, e.g. $p\to e^+\pi^0$. The strength of this operator scales like $M^{-2}$. Since all searches for this mode of proton decay have yielded negative results, it implies that $M\geq 10^{15}$ GeV or so. This kind of decay mode arises in Grand Unified Theories (GUTs) such as $SU(5)$ or $SO(10)$, and is the basis of the common lore that the scale of $B$-violation is very high.

However, $B$-violation could also manifest at low-energies in neutron-antineutron ($n-\bar{n}$)  oscillation~\cite{nnbar} as well as in di-nucleon decays, e.g. $np\to \pi^+\pi^0$ and $pp\to K^+K^+$~\cite{glashow, Arnold:2012sd} (for a review, see e.g.~\cite{Phillips:2014fgb}). These processes arise from generic higher dimensional effective operators, a typical one having the form  $u_Rd_Rd_Ru_Rd_Rd_R$  with $d=9$; the strength of this process, therefore, scales like $M^{-5}$. Due to the high power of $M$, current limits from  $n-\bar{n}$~\cite{ILL} and dinucleon decay searches~\cite{SK}  put a lower bound on the scale $M$ in the few TeV range depending on couplings in the theory. This makes it plausible that in a detailed ultra-violet (UV)-complete model which leads to this operator, the physics of $B$-violation can be tested in laboratory experiments. It is therefore important to study TeV-scale UV-complete models for $B$ violation and their usefulness for understanding the origin of matter in the Universe.

In this paper, we present a simple extension of the SM which provides a unified UV-complete theory for the TeV-scale $\Delta B=2$ operator and neutrino masses. The key features of this model are the addition of right-handed (RH) Majorana neutrinos $N_a$ $(a=1,2,3)$, a color-triplet scalar $\chi$ that connects them to the quark sector and a second inert Higgs doublet $\eta$. The Yukawa interactions of the color-triplet generate the effective $B$-violating operator $N_au_Rd_Rd_R$ which, in combination with the Majorana mass $M_{N_a}$ of the RH neutrinos, leads to the $d=9$, $\Delta B=2$ operator $u_Rd_Rd_Ru_Rd_Rd_R$. The same RH neutrinos, together with the second Higgs doublet, produce a small Majorana mass for the LH neutrinos at the one-loop level~\cite{ma}.  Their masses are of
order TeV and could therefore be searched for at colliders (for a review, see e.g.~\cite{Deppisch:2015qwa}).

Some of the basic  elements of this model are similar to the one presented in Ref.~\cite{babu1}, where a singlet fermion with Majorana mass was added to the SM together with a color triplet Higgs field; that helped to generate the $N_au_Rd_Rd_R$ operator and hence the $\Delta B=2$ operator for $n-\bar{n}$ oscillation.  However, the singlet fermion of Ref.~\cite{babu1} could not be identified with the RH neutrino since its Yukawa coupling to SM Higgs would lead to catastrophic proton decay, if the RH neutrinos have TeV-scale mass.  The new ingredient in the present paper  is to show that  by adding a second Higgs doublet and  a $Z_2$ symmetry which allows the second Higgs doublet to couple only  to the SM singlet fermions, one can now identify the singlet fermion as one of the RH neutrinos, which along with other RH neutrinos can play a role in generating the observed light neutrino masses and mixings. 
 Also the model can now be embedded in extended gauge groups such as $SU(2)_L\times U(1)_{I_{3R}}\times U(1)_{B-L}$ and possibly in higher groups such as $SO(10)$.

Another new result of this paper is a calculation of the matter-antimatter asymmetry in the model, which is now generated by a resonant baryogenesis mechanism with at least two quasi-degenerate RH neutrinos in the TeV range. The baryogenesis can either occur above or below the sphaleron decoupling temperature. In the latter case, this model provides a concrete realization of the post-sphaleron baryogenesis (PSB) scenario~\cite{psb}. Finally, the TeV-scale new particles in this model lead to interesting collider signals. 

Our model has the following low energy implications: (i)  it leads to $\Delta B=2$ processes such as $pp\to K^+K^+$ ~\cite{glashow} as well as $n-\bar{n}$ oscillation~\cite{Phillips:2014fgb} which have observable strengths (as in Ref.~\cite{babu1}); (ii) despite the second Higgs coupling to $N_a$, the presence of the unbroken  $Z_2$ symmetry prevents proton decay while allowing the previously mentioned $\Delta B=2$ processes;  (iii) light neutrino masses in this model arise via one loop seesaw diagram~\cite{ma}, which allows much larger Yukawa couplings for RH neutrinos than the canonical seesaw~\cite{seesaw}; (iv)  it leads to new collider signals with final states of type $pp\to \ell^+\ell^- + 6j$ and $pp\to 4j$ at the LHC. Finally, our model provides a testable mechanism for the origin of matter, which is qualitatively different from those discussed in Refs.~\cite{babu1,cheung, volkas, dutta, Monteux:2014hua, racker, kashiwa}.\footnote{After this work was completed, we  were informed of a related study~\cite{yue} on the implications of the $Nudd$ operator for baryogenesis and collider phenomenology.}

This paper is organized as follows: in Sec.~\ref{sec:2}, we give a description of the model;  Sec.~\ref{sec:3} focuses on the $\Delta B=2$ modes such as $pp\to K^+K^+$ and $n-\bar{n}$ oscillation in the model; Sec.~\ref{sec:4} discusses the neutrino mass generation at one loop level; Sec.~\ref{sec:5}  presents a calculation of the baryon asymmetry in the model via resonant baryogenesis; in Sec.~\ref{sec:6}, we comment on some collider signals and in Sec.~\ref{sec:7}, we conclude with a summary of the results.

\section{Description of the model}\label{sec:2}
We work within the SM gauge group $SU(3)_c\times SU(2)_L\times U(1)_Y$ and extend its particle content  by the addition of three RH neutrinos ($N_a$), an extra Higgs doublet ($\eta$) and an $SU(2)_L$-singlet, color-triplet scalar ($\chi$) with hypercharge $Y=+4/3$.\footnote{We have used the charge convention $Q=I_{3L}+\frac{Y}{2}$.} 
We denote the quark and lepton doublets of the SM by $Q_L^{\sf T}=(u_L, d_L)$ and $L^{\sf T}=(\nu_L, e_L)$, and singlets by $u_R, d_R, e_R$; the SM Higgs doublet field is denoted by $\phi$. We impose an additional $Z_2$-symmetry under which $Q_L, u_R, d_R, N_a$ and $\eta$ fields are taken to be odd, whereas $L, e_R, \phi$ and $\chi$ fields are taken to be even. 
The gauge- and $Z_2$-invariant interaction Lagrangian involving the leptons and the new fields $\chi$ and $N$ in the model is given by 
\begin{align}
{\cal L}_Y  \ =  & \  h_{\nu,a i}\overline{N}_a\eta L_i+ \frac{1}{2}M_{ab}N_a^{\sf T}C^{-1}N_b 
+\lambda_{aj}{N}^{\sf T}_a \overline{\chi}_\alpha u_{R, \alpha j}
\nonumber \\
& \  +\lambda^\prime_{ij}\epsilon^{\alpha\beta\gamma}\chi_\alpha d_{R,\beta i}d_{R,\gamma j}  +{\rm H.c.}
\label{lag}
\end{align}
where $\alpha,\beta,\gamma$ are color indices and $i, j, a, b$ are generation indices. Note that due to color anti-symmetry, the only non-zero $\lambda^\prime$'s are $\lambda^\prime_{12, 13, 23}$. 

From the $Z_2$ assignments, it is clear that proton and neutron are $Z_2$-odd whereas leptons and anti-leptons are $Z_2$-even. Thus, a single proton decay is forbidden  in this model, since it will always involve an odd number of  leptons in the final state. 
The  presence of the Majorana mass terms $M_{ab}$ in the Lagrangian~\eqref{lag} leads to the violation of lepton number ($L$) by two units. In conjunction with the effective $B$-violating operator $Nu_Rd_Rd_R$, this leads to an effective $\Delta B=2$ operator which gives rise to $n-\bar{n}$ oscillation,  as discussed below. 

The Higgs potential involving the SM Higgs doublet $\phi$ and the new doublet $\eta$ in the model is given by~\cite{ma}
\begin{align}
V(\phi,\eta)  & \ = \ -m_1^2 |\phi|^2 + m_2^2 |\eta|^2 + \lambda_1 |\phi|^4 + \lambda_2 |\eta|^4 \nonumber \\
& + \lambda_3 |\phi|^2 |\eta|^2 +\lambda_4 |\phi^\dagger \eta|^2 + \left[\frac{\lambda_5}{2}(\phi^\dagger \eta)^2 + {\rm H.c.}\right].
\label{pot}
\end{align}
The mass square of the SM Higgs field $\phi$ is negative so that it has a vacuum expectation value (VEV), i.e. $\langle \phi^0\rangle=v_{\rm wk}\equiv 174$ GeV, whereas the corresponding term for $\eta$ is positive, i.e. $\langle \eta\rangle=0$. The vanishing VEV of $\eta$  is radiatively stable due to the $Z_2$-symmetry. In the spirit of the model, we take $M_\eta$ to be in the TeV range and also choose $M_\eta \geq M_N$, since we assume the model to represent  TeV-scale physics. In the discussion below, we will take the following benchmark values for the masses: $M_\chi\sim 10 $ TeV and $M_N\sim 1$ TeV so that the model is testable in colliders as well as in low energy processes.

The fact that both $B$ and $L$ violation are realized at TeV-scale in this model imposes constraints on the various couplings in \eqref{lag}, in order to satisfy the flavor changing neutral current (FCNC) observations. For instance, in order to be consistent with the $K_L-K_S$ mass difference, we must have $\lambda^\prime_{13}\lambda^\prime_{23}\lesssim 10^{-3/2}$. Similarly, $B_d-\overline{B}_d$ and $B_s-\overline{B}_s$ oscillation observations lead to the constraints $|\lambda^\prime_{32}\lambda^\prime_{12}|\lesssim 10^{-1}$ and $|\lambda^\prime_{31}\lambda^\prime_{12}|\lesssim 10^{-1}$ respectively. These limits are not very strong because they arise only at the one-loop level. From the above discussion, we find that we could easily satisfy the FCNC constraints by assuming the conservative bounds $\lambda^\prime_{12}, \lambda^\prime_{32}\leq 10^{-2}$ for $M_\chi = 10$ TeV. The important point to note is that they still leave $\lambda^\prime_{13}$ unconstrained, so that its value could be of order $\sim 1$. This helps in obtaining an observable $n-\bar{n}$ oscillation as well as the correct baryon asymmetry, as we will see below.

\section{$B$ and $L$ violation} \label{sec:3}
The basic sources of $B$ violation are the last two terms in Eq.~\eqref{lag} involving  $\lambda$ and $\lambda^\prime$. To see this explicitly, note that the $\lambda^\prime$ coupling implies that the $\chi$-field has $B=-2/3$. The RH neutrino coupling  $\lambda$ together with the lepton Yukawa coupling $h_\nu$ would then conserve ``baryon number''. However, the term $\lambda_5(\phi^\dagger \eta)^2$ in the scalar potential ~\eqref{pot} would then break baryon number. Thus the model has $B$ violation even if the RH neutrinos did not have a Majorana mass in which case,  lepton number is still an unbroken symmetry. 

\begin{figure}[t!]
\centering
\includegraphics[width=7cm]{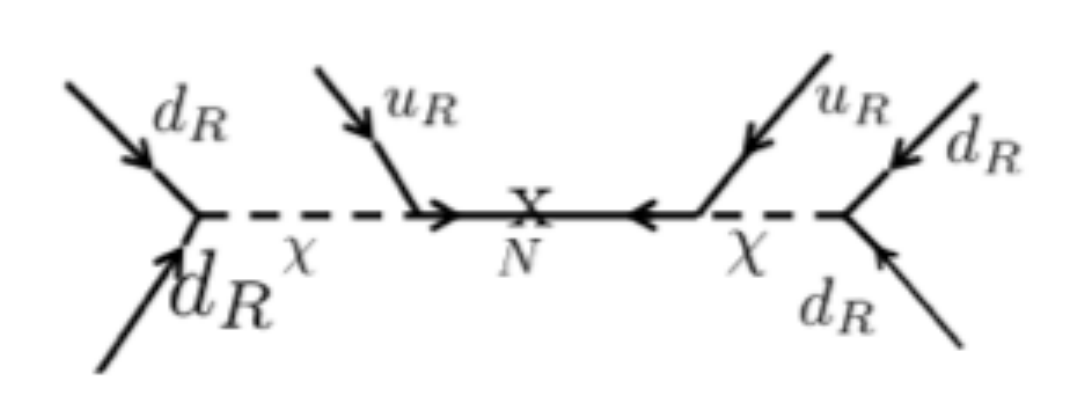}
\caption{Tree-level diagram for $\Delta B=2$ process induced by the effective operator~\eqref{eff1}.  All down-type quarks are denoted by the generic symbol $d_R$.}
\label{fig1}
\end{figure}

On the other hand, once the RH neutrinos $N_a$ have a Majorana mass, one could no more assign them a baryon number. In this case, both $B$ and $L$ would be broken by two units and the two breakings are connected to each other. In particular, the Majorana mass terms for $N_a$ would break both baryon and lepton number by two units. 
The starting effective $B$-violating operator in this case is  $N_au_Rd_Rd_R$~\cite{babu1} with a strength
\begin{eqnarray}
{\cal L}_I \ = \ \frac{\lambda_{ai}\lambda^\prime_{jk}}{M^2_\chi} N_a u_{R,i}d_{R,j}d_{R,k}+{\rm H.c.}
\label{eff1}
\end{eqnarray}
Combining this with the Majorana mass of the RH neutrinos, we get an effective $\Delta B=2$ operator at tree-level, as shown in Fig.~\ref{fig1}. 
Thus in this simple extension of SM, $\Delta L=2$ implies $\Delta B=2$.  

As noted below Eq.~\eqref{lag}, due to color anti-symmetry of the coupling $\lambda'$, the two down-type quarks coupling to $\chi$ must involve different families. Hence, the leading tree-level $\Delta B=2$ operator must change strange or bottom quantum number by two units.
For example, the strength of the effective operator with $\Delta s=2$ is given by\footnote{There are similar operators involving $b_R$. However, these ones are not of interest for our purpose, since they do not lead to di-proton decay.}
\begin{align}
{\cal L}_{\Delta B=2}  \ = & \ \frac{(\lambda_{a1}\lambda^\prime_{12})^2}{16\pi^2 M^4_{\chi} M_{N_a}} \epsilon^{ijk}\epsilon^{lmn}(u^{\sf T}_{R,i}C^{-1}u_{R,l}) \nonumber \\
& 
(d^{\sf T}_{R,j}C^{-1}s_{R,k}) (d^{\sf T}_{R,m}C^{-1}s_{R,n})+~ {\rm H.c.},
\end{align}
where $i,j,k,l,m,n$ are color indices. This $\Delta s=2$,  $\Delta B=2$ operator leads to the di-proton decay $pp\to K^+K^+$, whose lifetime is constrained to be $\tau_{pp\to KK}\geq  1.7\times 10^{32}$ yr~\cite{SK}. In order to translate this bound into bounds on couplings,  we need to go from six quarks to two protons. This transition would involve QCD dressing and has been discussed in the context of MIT bag models~\cite{shrock} as well as lattice models for QCD~\cite{buchoff}. 
Using the same dressing factor $\sim 10^{-5}$, we find that for $pp\to KK$ decay rate to be consistent with the current experimental limit~\cite{SK}, we must have 
$\lambda^\prime_{12}\lambda_{a1}\lesssim 10^{-4}$. Thus, we can choose $\lambda^\prime_{12}\leq 10^{-4}$ to satisfy the di-proton decay constraint,  while keeping $\lambda_{a1}\sim 1$ which helps for the purpose of baryogenesis, as discussed in the next section.

Note that as far as the $\Delta B=1$ operator is concerned, there is one operator of the form $\eta Lu_Rd_Rd_R$ induced by the exchange of $\chi$ and $N_a$  fields. This could have been seen from $Z_2$ invariance of the model: it implies also that since $L$ is even under this symmetry, the only way it can combine with the $Z_2$ odd $u_Rd_Rd_R$ operator is, when it appears together with the $Z_2$ odd field $\eta$. Since $\eta$ does not have a VEV, this operator cannot induce proton decay. 
Note however that the $\eta$ field could be pair-produced in colliders via SM $Z$ or photon exchange and would lead to $B$-violating final states, as discussed later.

\begin{figure}[t!]
\centering
\includegraphics[width=7cm]{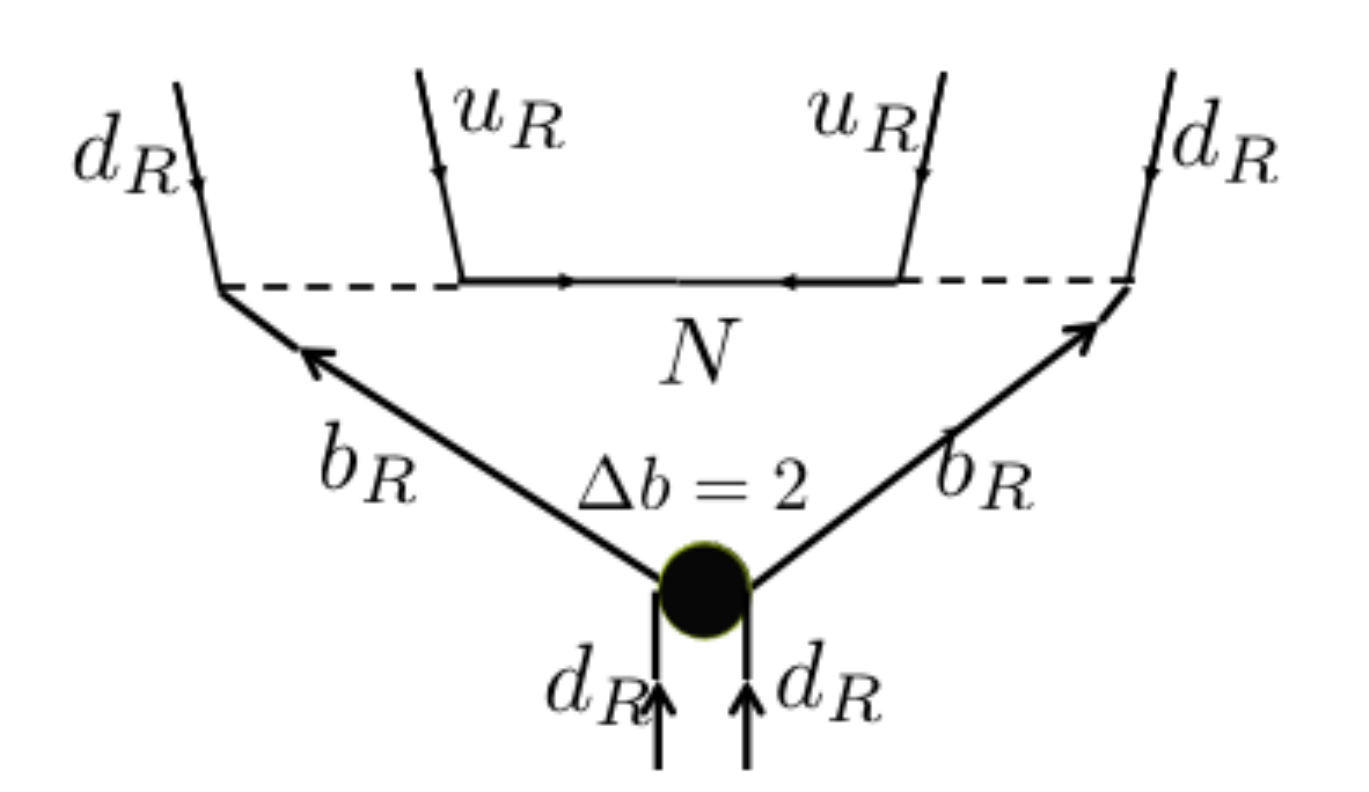}
\caption{One loop diagram for $n-\bar{n}$ oscillation using $\Delta b=2$ operator.}\label{fig2}
\label{nnbarfigure}
\end{figure}

To get $n-\bar{n}$ oscillation in this model, one has to convert two strange or bottom quarks  to two down quarks.  This will need a $\Delta s=2$ or $\Delta b=2$ effective interaction. Due to the constraints from $pp\to KK$ life time, the dominant contribution comes from the $\Delta b=2$ operator, which can be parameterized as $(\bar{d}_R\gamma^\mu b_R)^2/\Lambda^2$. In combination with the $\Delta B=2$ operator shown in Fig.~\ref{fig1}, it gives rise to $n-\bar{n}$ oscillation at one-loop level, as shown in Fig.~\ref{nnbarfigure}. The strength of this $n-\bar{n}$ operator is given by 
\begin{eqnarray}
{G}_{n-\bar{n}} \ \simeq \ \frac{(\lambda_{a1}\lambda^\prime_{13})^2M_{N_a}}{16\pi^2 M^4_{\chi}\Lambda^2}\ln\left(\frac{\Lambda^2}{M^2_{N_a}}\right) \, .
\end{eqnarray}
Using $\Lambda\sim 10^6$ GeV to satisfy the constraints of $B_d-\overline{B}_d$ mass difference, we find that $\tau_{n-\bar{n}}\geq 3\times 10^{8}$ sec, as required by the current limits~\cite{ILL},  if  $(\lambda_{a1}\lambda^\prime_{13})\leq 10^{-1}$. Note that both these couplings are unsuppressed by FCNC constraints, and therefore, can be of order $\sim 1$, thus giving rise to a large $n-\bar{n}$ amplitude, which is in the observable range of currently planned experiments. 

\begin{figure}[t!]
\centering
\includegraphics[width=7cm]{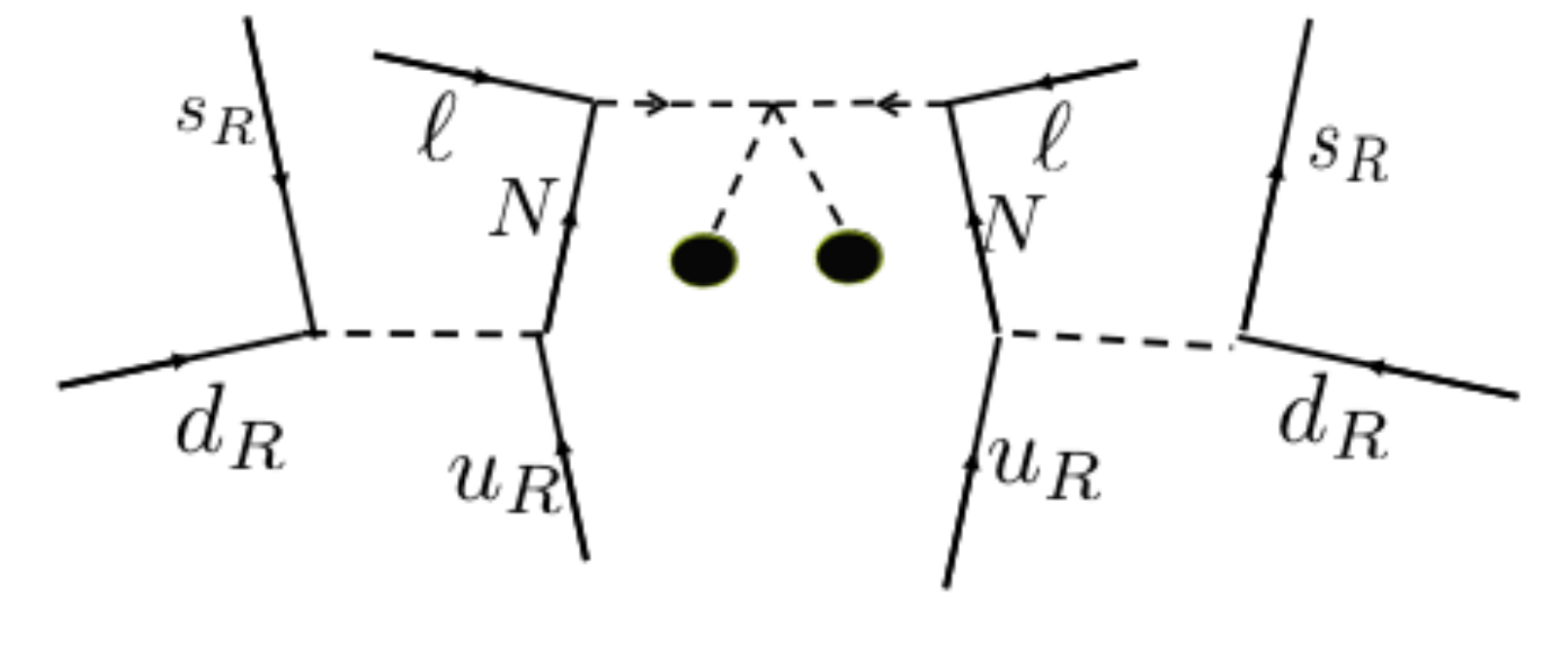}
\caption{Tree-level diagram for the process $pp\to KK\bar{\nu}\bar{\nu}$. }
\label{fig3}
\end{figure}

Due to the fact that the $N_a$'s are identified with RH neutrinos, this induces a tree level leptonic $B$-violating process via the diagram shown in Fig.~\ref{fig3}. This leads to the $\Delta B=2,~\Delta L=2$ process $pp\to K^+K^+\bar{\nu}\bar{\nu}$. However, the smallness of the $\lambda^\prime_{12}$ coupling as assumed above is enough to suppress this process to an unobservable level.

\section{Neutrino Mass} \label{sec:4}
\begin{figure}[t!]
\centering
\includegraphics[width=7cm]{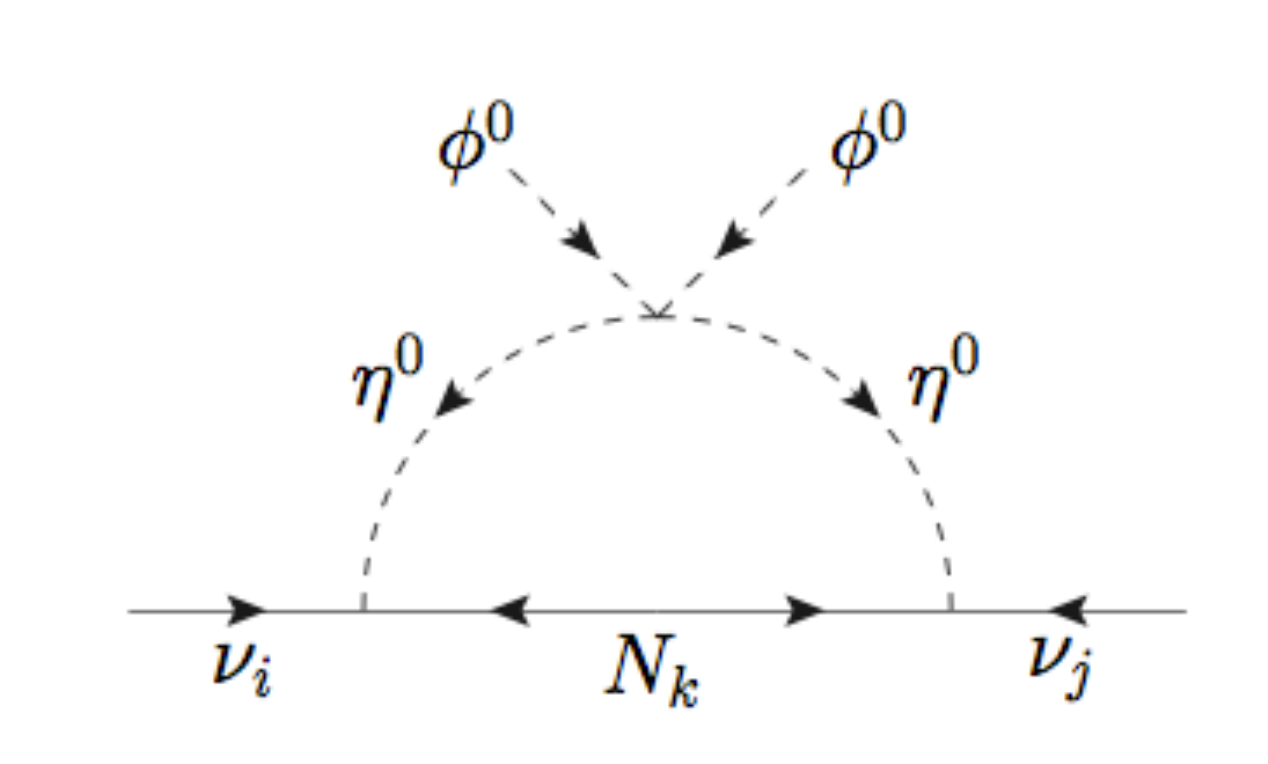}
\caption{One-loop graph for radiative seesaw.}
\label{fig4}
\end{figure}

To understand the origin of neutrino mass in this model, we first note that the $Z_2$ symmetry forbids the usual Dirac mass term $\overline L \phi N$.  The leading contribution to neutrino mass comes from a one-loop graph involving $\eta$ and $N_a$, as suggested in Ref.~\cite{ma} and shown in  Fig.~\ref{fig4}.
We assume that masses of the $\eta$ Higgs components (denoted generically by $M_\eta$) are much larger than the masses of the RH neutrinos $M_a$. 
In this limit, the light neutrino masses are given by~\cite{ma}
\begin{eqnarray}
({\cal M}_\nu)_{ij} \ \simeq \ \frac{\lambda_5v^2_{\rm wk}}{8\pi^2 M^2_\eta} h_{\nu, ai}h_{\nu, aj}M_{N_a} \, .
\label{eq:ma}
\end{eqnarray}
This implies that we can get the desired neutrino masses in the sub-eV range for TeV-scale RH neutrinos if $M_\eta \sim 10$ TeV and Yukawa couplings $h\sim 10^{-7/2}-10^{-4}$. These values of the couplings seem reasonable, as they are comparable to some of the lepton Yukawa couplings in the  SM. We will see below that the assumption of $M_\eta \gg M_{N_a}$ is useful in understanding baryogenesis without any simultaneous leptogenesis from $N$ decays.
Moreover, to facilitate resonant baryogenesis, we must have at least two quasi-degenerate RH neutrinos, so that there is a direct relationship between the neutrino mass and Yukawa couplings: $
{\cal M}_\nu\simeq \alpha_{\rm loop} h_\nu h^{\sf T}_\nu$, where $\alpha_{\rm loop}$ is the one-loop factor in Eq.~\eqref{eq:ma}, i.e. $\alpha_{\rm loop}\sim \lambda_5v^2_{\rm wk}M_{N_a}/8\pi^2 M^2_\eta$.

\section{Resonant Baryogenesis} \label{sec:5}
\begin{figure}[t!]
\centering
\includegraphics[width=7cm]{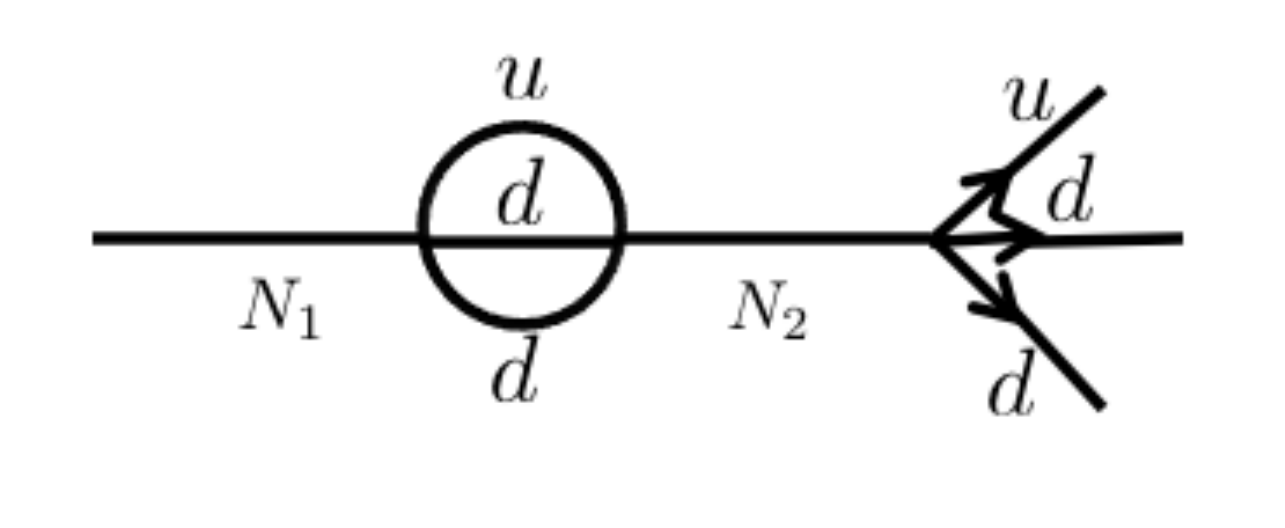}
\caption{RH neutrino self-energy diagram responsible for resonant baryogenesis in our model. }
\label{fig5}
\end{figure}

The baryon asymmetry in this model is directly obtained from the $B$-violating out-of-equilibrium decay of the RH neutrinos. 
If we assume that $M_\eta, M_\chi \gg M_{N_a}$, 
the only decay mode of $N_a$ in the early Universe that is relevant for baryogenesis is the 3-body decay $N\to u_Rd_Rd_R$ via $\chi$ exchange (cf. Fig.~\ref{fig1}), governed by the effective interaction \eqref{eff1}. A non-zero $CP$-asymmetry can be generated due to the interference of the tree-level decay graph with the one-loop graphs containing an absorptive part. As in the case of leptogenesis, there are two contributions, namely, $\varepsilon'$-type $CP$-asymmetry which is dominant in the hierarchical case $M_{N_{2,3}}\geq M_{N_1}$~\cite{Buchmuller:2004nz}, and $\varepsilon$-type $CP$-asymmetry which is dominant in the quasi-degenerate case~\cite{resonant}. Here we consider the second case (cf. Fig.~\ref{fig5}), where the $CP$-asymmetry is resonantly enhanced by the RH neutrino self-energy effects~\cite{resonant}. This idea is qualitatively different from the mechanisms studied in Refs.~\cite{babu1,cheung, dutta, volkas, Monteux:2014hua, racker, kashiwa}. Since the RH neutrino decay directly produces a baryon asymmetry in our model without relying on the sphaleron transitions, we call this the {\em resonant baryogenesis} scenario.\footnote{For a discussion of baryogenesis in the hierarchical case, see~\cite{cheung}. } 

As the Universe evolves, the new heavy particles $\eta$ and $\chi$ disappear via decay and annihilation and only the quasi-degenerate RH neutrinos remain in equilibrium down to the TeV temperature. Their out-of-equilibrium decay to three quarks is the dominant source term for generating the baryon asymmetry. 
At tree-level, the total decay rate for $N_a\to u_Rd_Rd_R+{\rm C.c.}$ is given by 
\begin{align}
\Gamma_{N_a} & \ \simeq \ \frac{1}{512\pi^3}\sum_{i,j,k} |\lambda_{ai}\lambda^{\prime}_{jk}|^2 \frac{M_{N_a}^5}{M_\chi^4} \; ,
\label{decay}
\end{align} 
in the limit $M_\chi \gg M_{N_a}$. 
For two quasi-degenerate RH neutrinos, the interference between the tree-level graph with the one-loop self-energy graph (cf. Fig.~\ref{fig5}) gives the dominant contribution to the $CP$-asymmetry: 
\begin{align}
\varepsilon  \ \simeq \ & \frac{1}{3072\pi^3}\frac{M^4_N M_{N_1}M_{N_2} (M_{N_1}^2-M_{N_2}^2)}{M^4_\chi [(M_{N_1}^2-M_{N_2}^2)^2+M_N^2\Gamma_N^2]}\nonumber \\ 
\times & \frac{\sum_{i,j,k,l,m,n}{\rm Im}[(\lambda_{1i}\lambda_{jk}')(\lambda_{1l}\lambda_{mn}')^{\ast}(\lambda_{2l}\lambda_{mn}'
)(\lambda_{2i}\lambda_{jk}')^{\ast}]
}{\sum_{i,j,k} (|\lambda_{1i}\lambda^\prime_{jk}|^2+|\lambda_{2i}\lambda^\prime_{jk}|^2)}  \, ,
\label{cp}
\end{align}
where $M_N$ is the average mass and $\Gamma_N$ is the average width. The $\varepsilon$-type $CP$-asymmetry gets resonantly enhanced when $\Delta M_N\equiv |M_{N_1}-M_{N_2}|\sim \Gamma_N/2 \ll M_N$. Such a mass degeneracy could arise naturally, for example, due to a symmetry under which $N_1$ and $N_2$ have opposite quantum numbers. 
Clearly, for a given set of couplings $\lambda_{ai},\lambda'_{ai}$ satisfying all the experimental constraints, $\Delta M_N$ can be arranged such that the $CP$ asymmetry is adequate to explain the observed baryon asymmetry. 
For example, with a 1\% level degeneracy i.e. $\Delta M_N/M_{N}\sim 0.01$ and ${\cal O}(1)$ couplings, we can get $\varepsilon \sim 10^{-4}$ from Eq.~\eqref{cp}. Due to FCNC and diproton decay constraints, the dominant contribution to Eq.~\eqref{cp} comes from  $bdu, bdc,bdt$-quark intermediate states in Fig.~\ref{fig5}.

For a given set of model parameters, we should compare the decay rate~\eqref{decay} with the Hubble expansion rate 
\begin{align}
H(T)  = 1.66\sqrt{g_*}\frac{T^2}{M_{\rm Pl}}\; , 
\end{align}
where $T$ is the temperature, $g_*$ is the effective relativistic degrees of freedom and $M_{\rm Pl}=1.2\times 10^{19}$ GeV is the Planck mass. Defining a quantity $K=\frac{\Gamma_N}{\zeta(3)H_N}$, where $z\equiv M_N/T$, $H_N\equiv H(z=1)$ and $\zeta(x)$ is the Riemann zeta function, we find that for $K\gtrsim 3$, the system is in the strong washout regime~\cite{Buchmuller:2004nz}, where the final baryon asymmetry generated around $z=1$ is insensitive to any pre-existing asymmetry at $z\ll 1$. For ${\cal O}(1)$ couplings and $M_\chi = 10$ TeV, this is  satisfied in our model 
 for $M_N\gtrsim 10$ GeV, and for lower values of $M_\chi$, it can be achieved for lower values of $M_N$.   

For $M_N\gtrsim T_c\simeq 150$ GeV, where $T_c$ is the critical temperature below which the electroweak sphaleron processes become ineffective~\cite{Cline:1993bd}, the produced baryon asymmetry will get somewhat diluted due to the sphaleron effects. It should be noted here that since the baryon asymmetry in our model is produced solely in the RH-quark sector, the sphaleron effects do not directly act on it, but part of the asymmetry transferred to the LH-quark sector will get diluted due to SM Yukawa and QCD interactions. There is an additional entropy dilution effect due to standard photon production from the epoch of $T_c$ to the recombination epoch $T_0$, at which the baryon-to-photon ratio $\eta_{\Delta B} \equiv (n_B-n_{\overline{B}})/n_\gamma$ is measured. Putting these effects together, we find that the net dilution factor in our model is $d\simeq 2.4\times 10^{-2}$. Thus, we need to produce a primordial baryon asymmetry of $\eta_{\Delta B}(T_c)\simeq 2.5\times 10^{-8}$ in order to be consistent with the observed value of $\eta_{\Delta B}^{\rm obs} =(6.101^{+0.086}_{-0.081})\times 10^{-10}$~\cite{Planck:2015xua}.

For $M_N\lesssim T_c$, the produced baryon asymmetry is not affected by the sphaleron processes, and hence, the only dilution effect is due to the entropy production: $d\simeq g_s(T_0)/g_s(T=M_N)$, where $g_s(T_0)=3.91$ is the effective degrees of freedom corresponding to the entropy density at recombination and $g_s(T=M_N)$ depends on the RH neutrino mass scale $M_N$.  This scenario provides a concrete realization of  the PSB mechanism~\cite{psb}.

For the thermodynamic evolution of the generated baryon asymmetry, we should also take into account the washout effects, mainly due to the scattering processes $N_a\bar{u}_R\to d_Rd_R$ and $N_a\bar{d}_R\to u_Rd_R$, as well as the inverse decay $u_Rd_Rd_R\to N_a$, all mediated by $\chi$.\footnote{Other scattering processes involving $\chi$ in the initial state will be Boltzmann-suppressed at $T\lesssim M_N$ for $M_\chi\gg M_N$.}  
The thermally-averaged decay, inverse decay and scattering rates are respectively given by 
\begin{align}
\gamma_{D_a} \  & = \ \frac{TM_{N_a}^2}{\pi^2} \Gamma_{N_a} K_1(M_{N_a}/T) \; , \label{dec}\\
\gamma_{I_a} \ & = \ \frac{1}{2}\gamma_{D_a} \frac{\eta_{N_a}^{\rm eq}}{\eta_B^{\rm eq}} \; , \label{inv} \\
\gamma_{S_a} \ & = \frac{T}{64\pi^4} \int_{m_{N_a}^2}^{\infty} ds\: \sqrt s \: \hat{\sigma}_a(s)\: K_1(\sqrt s/T)  \; , \label{scat}
\end{align}
where $K_n(x)$ is the $n$-th order modified Bessel function of the second kind, $\eta^{\rm eq}_{N_a} = \frac{z^2K_2(z)}{2\zeta(3)}$, $\eta_B^{\rm eq} = \frac{3}{4}$, and $\hat{\sigma}$ is the reduced cross section~\cite{Luty:1992un}:
\begin{align}
\hat{\sigma}(s) \ = \ \frac{1}{8\pi s}\int_{t_{\rm min}}^{t_{\rm max}} dt \sum_{\rm spins} |{\cal M}|^2 \; ,
\label{reduced} 
\end{align}
where $s, t$ are the usual Mandelstam variables and ${\cal M}$ is the $2\leftrightarrow 2$ scattering matrix element. 
For both the processes $N_a\bar{u}_R\to d_Rd_R$ and $N_a\bar{d}_R\to u_Rd_R$, we have $t_{\rm min}=M_{N_a}^2-s$ and $t_{\rm max}=0$. Using Eq.~\eqref{reduced}, we thus obtain respectively for $N_a\bar{u}_R\to d_Rd_R$ and $N_a\bar{d}_R\to u_Rd_R$,  
\begin{align}
\hat{\sigma}_{a1}(s) \  = \ & \frac{3}{2\pi} \sum_{i,j,k} |\lambda_{ai}\lambda^{\prime}_{jk}|^2 \frac{(s-M_{N_a}^2)^2}{[(s-M_\chi^2)^2+M_\chi^2\Gamma_\chi^2]} \; ,\\
\hat{\sigma}_{a2}(s) \  = \ & \frac{3}{2\pi s} \sum_{i,j,k} |\lambda_{ai}\lambda^{\prime}_{jk}|^2 \left[\frac{(s-M_{N_a}^2)(s+2M_\chi^2)}{s+M_\chi^2}\right. \nonumber \\
& \qquad \left. -\: 2M_\chi^2\log\bigg(1+\frac{s}{M_\chi^2}\bigg)\right] \; .
\end{align}
The final baryon asymmetry is obtained by solving the following coupled Boltzmann equations: 
\begin{align}
\frac{d\eta_{N_a}}{dz} \ & = \ -\left(\frac{\eta_{N_a}}{\eta^{\rm eq}_{N_a}}-1\right)(D_a+S_a),\label{be1}\\
\frac{d\eta_{\Delta B}}{dz} \ & = \  \sum_a \left(\frac{\eta_{N_a}}{\eta^{\rm eq}_{N_a}}-1\right)\varepsilon D_a - \eta_{\Delta B} \sum_a W_a \, , \label{be2}
\end{align}
where $\eta_{\Delta B}$ denotes the {\em total} baryon asymmetry, i.e. summed over all quark flavors. Including the flavor off-diagonal effects could lead to an enhanced asymmetry, depending on the model parameters~\cite{flav}, but we do not discuss it here for simplicity. The various reaction rates in Eqs.~\eqref{be1} and \eqref{be2} are defined using Eqs.~\eqref{dec}-\eqref{scat}: 
\begin{align}
D_a & \ = \ \frac{z}{H_Nn_\gamma}\gamma_{D_a}\; , \\
S_a & \ = \ \frac{z}{H_Nn_\gamma}2(\gamma_{S_{a1}}+\gamma_{S_{a2}})\; , \\
W_a & \ = \ \frac{z}{H_Nn_\gamma}(\gamma_{I_a}+2\gamma_{S_{a1}}+2\gamma_{S_{a2}})\; . 
\end{align} 

\begin{figure}[t!]
\centering
\includegraphics[width=8cm]{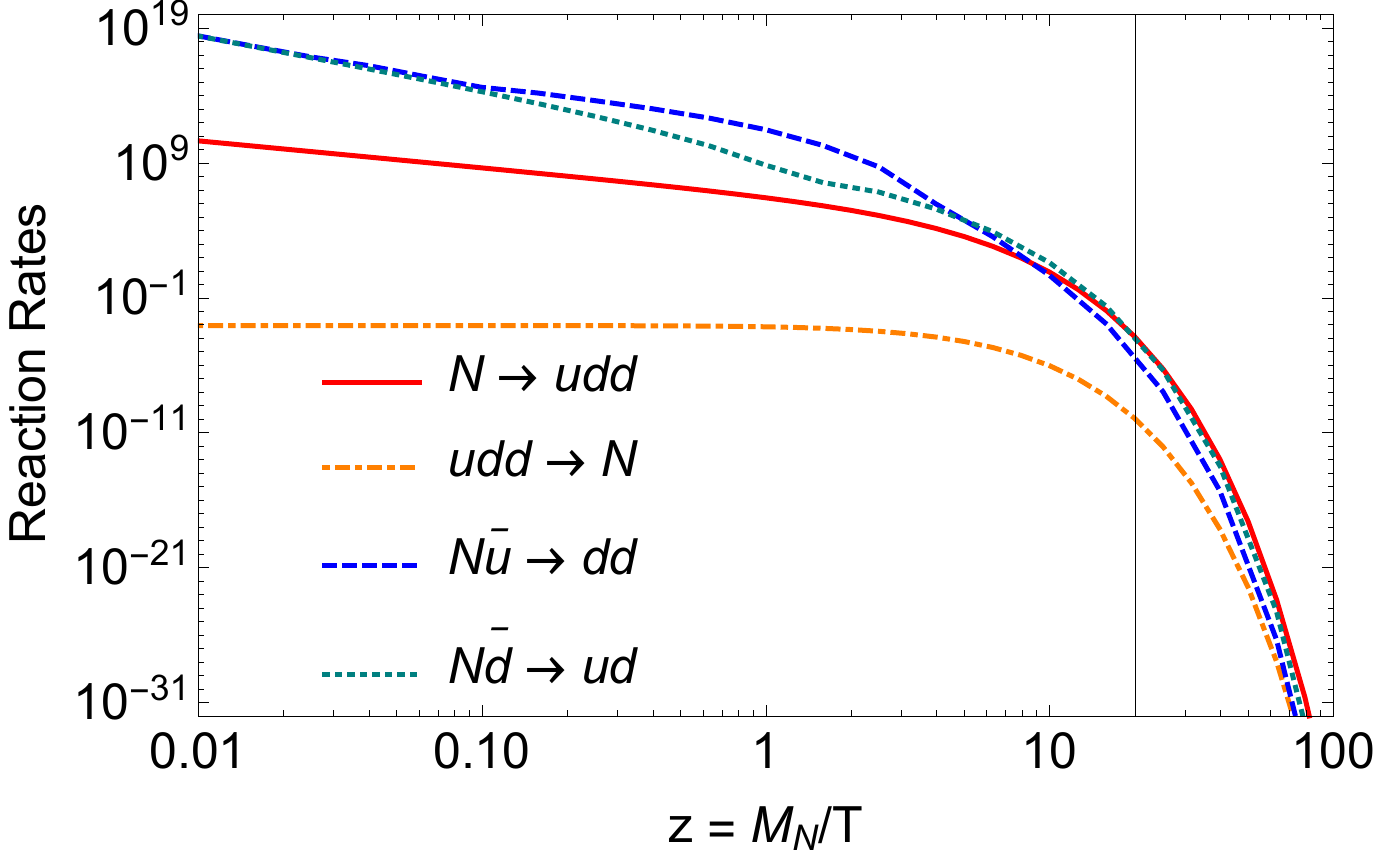}
\caption{Comparison of the decay, inverse decay and scattering rates. The vertical line is for $z_c=M_N/T_c$. }
\label{fig7}
\end{figure}

In  Fig.~\ref{fig7}, we compare the reaction rates defined in Eqs.~\eqref{dec}-\eqref{scat} for a benchmark case with ${\cal O}(1)$ couplings, $M_N=3$ TeV and $M_\chi=10$ TeV. We find that the 3-body inverse decay rate is much smaller than the decay rate, and moreover, the decay rate is dominant over the scattering rates for $z\sim z_c$, as required for successful baryogenesis in the strong washout regime. Utilizing the resonant enhancement of the $CP$-asymmetry~\eqref{cp}, we find that the required baryon asymmetry can be generated above $T_c$ for $M_N\gtrsim 1$ TeV, whereas for the post-sphaleron case, lower values of $M_N$ are also possible. 


\section{Comments}  \label{sec:6}
Now we make several comments on the phenomenological implications of our  model, including a brief description of some novel collider signals.

\noindent (a)  In usual type I seesaw models, the RH neutrino couples to the SM Higgs, which acquires a VEV. This leads to mixing of $N$'s with light neutrinos  (for a review of the phenomenology of this class of models, see e.g.~\cite{Alekhin:2015byh}). Thus $N$ production in colliders occurs via the $\nu-N$ mixing. However, in our model, there is no $\nu-N$ mixing, so the RH neutrinos can only be produced at colliders from the $\eta^\pm \to \ell^\pm N$ decay, where the inert doublet $\eta$ can be produced in pairs at the LHC in a Drell-Yan process. In the usual inert doublet model~\cite{ma}, if $M_N \leq M_\eta$, as we assume here, the final state $N$ would go undetected as a missing energy. However, in our model, since $N\to udb$ is allowed, we can have $\eta\to \ell jjb$ if $|M_{\eta^+}-M_{\eta^0}| \leq M_W$. This condition is required to satisfy the bounds from $T$ parameter in this model~\cite{pdg}. Thus in this parameter range, the LHC signal for this model would be $\ell^\pm \ell^\mp 4jb{b}$. For lower mass $\eta$'s, $e^+e^-$ colliders may be ideal to search for this new signal~\cite{ILC}. Note that this is different from the $\ell^\pm\ell^\pm jj$ signal of the canonical type I seesaw~\cite{KS}. Another interesting collider signal is: $pp\to \bar{q}N\to 4j$ + no $\slashed{E}_T$. A detailed analysis of these collider signals and the relevant QCD background will be presented elsewhere. 

\noindent (b) In contrast with the inert doublet model, in our case, the $\eta^0$ field is unstable and therefore does not play the role of dark matter.

\noindent (c) The contribution to neutrinoless double beta decay in this model comes only from the light neutrinos~\cite{racah} and there are no heavy particle contributions.

\noindent (d) This model could be embedded into an $SU(2)_L\times U(1)_{I_{3R}}\times U(1)_{B-L}$ theory above the scale of $M_\chi$. This is the next minimal anomaly free gauge extension of the model. This embedding and its eventual GUT embedding in $SO(10)$ is currently under investigation.

\section{Conclusion}  \label{sec:7}
We have presented a simple TeV-scale model for both $B$ and $L$ violation which can provide an understanding of neutrino masses without fine tuning of parameters and also a resonant baryogenesis mechanism for explaining the origin of matter. A supersymmetric version of this model is also known to provide a candidate for dark matter of the universe~\cite{babu1}. The model is testable via its prediction of the $\Delta B=2$ processes such as $pp\to K^+K^+$ and $n-\bar{n}$ oscillations. The latter might be  observable with current and future experimental facilities. A novel six jet plus dilepton signal of this model can be searched for at the LHC.

\section*{Acknowledgements}
We are grateful to Hooman Davoudiasl and Yue Zhang for helpful discussions and comments, and in particular, for informing us about their paper on a related subject~\cite{yue}. 
The work of P.S.B.D. is supported by the Lancaster-Manchester-Sheffield Consortium for
Fundamental Physics under STFC grant ST/L000520/1. The work of R.N.M. is supported
in part by the National Science Foundation Grant No. PHY-1315155. P.S.B.D. acknowledges the local hospitality provided by the Brookhaven National Laboratory and the University of Massachusetts,   Amherst, where part of this work was completed. 


\end{document}